\documentclass[useAMS,usenatbib,longnamesfirst]{mn2e}

\usepackage{epsfig}
\bibpunct{(}{)}{;}{a}{}{,}
\newcommand\jcd{Christensen-Dalsgaard}
\newcommand\ea{et al.}

\title[Helium abundance in stellar envelopes]{Asteroseismic determination
of helium abundance in stellar envelopes}
\author[S. Basu et al.]{Sarbani Basu$^{1}$\thanks{E-mail:
basu@astro.yale.edu}, Anwesh Mazumdar$^{2}$, H. M. Antia$^{3}$ and Pierre
Demarque$^{1}$\\
$^{1}$Astronomy Department, Yale University, P. O. Box 208101,
New Haven CT 06520-8101, U. S. A.\\
$^{2}$LESIA, Observatoire de Paris, 5 Place Jules Janssen, Meudon 92190, France\\
$^{3}$Tata Institute of Fundamental Research, Homi Bhabha Road,
Mumbai 400005, India}
\begin{document}

\date{Accepted . Received ; in original form }


\maketitle

\label{firstpage}

\begin{abstract}
Intermediate degree modes of the solar oscillations have previously been
used to determine the solar helium abundance to a high degree of
precision. However, we cannot expect to observe such modes in other
stars. In this work we investigate whether low degree modes that should
be available from space-based asteroseismology missions can be used to
determine the helium abundance, $Y$,  in stellar envelopes with sufficient
precision. We find that the oscillatory signal in the frequencies caused
by the depression in $\Gamma_1$ in the second helium ionisation zone can
be used  to determine the envelope helium abundance of low mass main
sequence stars. For frequency errors of 1 part in $10^4$, we expect
errors $\sigma_Y$ in the estimated helium abundance to range from $0.03$
for 0.8M$_\odot$ stars to  $0.01$ for 1.2M$_\odot$ stars.  The task is
more complicated in evolved stars, such as subgiants, but is still
feasible if the relative errors in the frequencies are less than
$10^{-4}$.
\end{abstract}


\section{Introduction}
\label{sec:intro}

The helium abundance of stars is often used to extrapolate back to obtain
estimates of the primordial helium abundance which is an important test of
cosmological models. The helium content of the oldest stars cannot
however, be measured directly --- these are low mass stars, and their
photospheres  are not hot enough 
to excite helium lines that can be used to determine the helium
abundance spectroscopically. Thus the helium content of low-mass stars
has to be derived from the evolution of heavier elements via stellar models. 
The evolution of stars however depend crucially on the helium
content itself, making the  knowledge of the helium abundance in stars
very important.
The estimated ages of globular clusters, which provides an important
constraint on cosmological models, are known to depend sensitively on
the efficiency of helium diffusion in the envelope \citep{ddk90,pv91,dd91,cha92}.

\begin{figure}
  \begin{center}
    \leavevmode
  \centerline{\epsfig{file=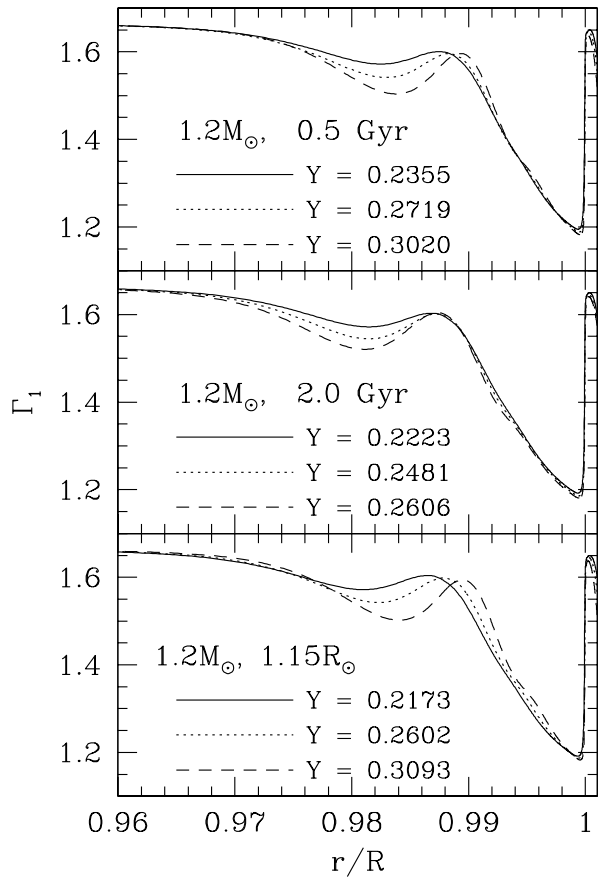,width=7.9cm}}
  \end{center}
  \caption{The adiabatic index, $\Gamma_1$, as a function
of fractional radius for 1.2~$M_\odot$ models. The upper two panels
are models with different $Y$ evolved to the age indicated in the
panels. The bottom panel shows $\Gamma_1$ for 1.2~$M_\odot$ models evolved
till the surface is at a radius of $1.15R_\odot$.
The lines of different types represent models with different
helium abundances as marked in each panel.}
\label{fig:gam1}
\end{figure}

The helium abundance of the solar envelope  has been successfully determined
using helioseismic data \citep[][ etc.]{gou84,dap91,ba95}
These seismic determinations were based on the fact that
the ionisation of hydrogen and helium causes a distinct, localised depression
in the adiabatic index, $\Gamma_1$, in the near-surface layers
of the Sun. The first helium ionisation zone  is not very useful in making
seismic inferences --- it overlaps with the
hydrogen ionisation zone and is located very close to the stellar
surface where there are significant uncertainties in seismic
inversions. 
 The second helium ionisation
zone which is located below the highly superadiabatic layer of the
convection zone is not particularly sensitive to surface effects and hence is
useful in determining the helium abundance.
The depression of $\Gamma_1$
in this region can be directly related to the
abundance of helium; this depression increases
with increase in the helium abundance (see Fig.~\ref{fig:gam1}).
The depression in $\Gamma_1$ causes a localised depression in
the derivative of the sound speed in the region, and hence affects all 
acoustic modes that travel through that region.
Thus using  helioseismic inversion techniques it
is possible to determine
the sound speed in the solar interior which in turn could be calibrated
to obtain the helium abundance in the solar envelope.
All studies to determine the helium abundance in the Sun used
intermediate degree modes to measure the sound speed in the HeII ionisation zone.

A number of current and proposed space based asteroseismic missions
like MOST \citep{wal03}, COROT \citep{baglin03}, MONS \citep{kcb03}
and Eddington \citep{rf03}
are expected to measure the oscillation frequencies of  several
stars with sufficient accuracy to enable a more detailed study of stellar
structure than has been possible so far.
Ground based observations have already started measuring the
frequencies of many stars despite the difficulties of 
such measurements \citep{bc02}.
Unfortunately, none of these observations can or will be able
to observe the intermediate degree modes  of oscillation that have been
so useful in solar studies. These missions can
at most, hope to determine the frequencies of modes
with degree $\ell\le 3$. Therefore, it is important to devise
a practical way for determining the helium content of stellar
envelopes using only low degree modes.

Any localised feature in the sound speed inside a star, 
such as that caused by 
the change in the temperature
gradient at the base of the convection zone, introduces an oscillatory
term in
frequencies as a function of radial order $n$, that is proportional to
\begin{equation}
 \sin(2\tau_m\omega_{n,\ell} + \phi), \qquad \tau_m=\int^R_{r_m} {dr\over c}\;,
\label{eq:osc}
\end{equation}
\citep[e.g.,][]{gou90} where $\tau_m$ is the acoustic depth of the localised
feature, $c$ the speed of sound, $r_m$ the radial distance where the
feature is located; and $\omega_{n,\ell}$ the angular frequency of a
mode with radial order $n$ and degree $\ell$,
$\phi$ is a phase.  
This oscillatory signature has been extensively studied for the
Sun in order to determine the extent of overshoot below the solar
convection zone \citep{mct94,ban94}.
The HeII ionisation zone too gives rise to a similar oscillatory
signal, which can 
be extracted from  frequencies of low degree modes \citep[see for
example,][]{rv01}. The present work is motivated by the possibility of
using this signal in low degree seismic data from distant stars to study
the properties of the HeII ionisation zone.

The depression in $\Gamma_1$ in the HeII ionisation zone is not a very sharp feature
and has a finite width. However, one can show that it still
gives rise to an oscillatory signal in the frequencies
and that the 
amplitude of the oscillatory term caused by the
depression in $\Gamma_1$ depends on the amount of helium present at the
location of ionisation.
In principle, if this amplitude is measured, it can be compared against 
models to estimate the helium abundance in the stellar envelope.
\citet{mt98} have studied the oscillatory signal
arising from the HeII ionisation zone in stellar models.
They have shown that because of the
finite width of this zone, the amplitude of oscillatory
signal in frequencies is modulated by an additional factor
\begin{equation}
\sin^2(\omega\beta)/(\omega\beta),
\label{eq:factor}
\end{equation}
where $\beta$ is half the acoustic
thickness of the dip in $\Gamma_1$. Since $\beta$ is rather small,
of the order of 100~s, this factor provides a slow modulation in
amplitude of the oscillatory term. \citet{mig03} have also
examined the oscillatory signal due to HeII ionisation zone and
pointed out the possibility of using it to determine the helium
abundance.

\citet{bad03} examined several stellar models
to study the possibility of determining helium abundance from the
amplitude of the oscillatory signal. They calculated the amplitudes
using the second differences of the frequencies.
In this work we extend that study to a much larger sample of models
to explore possible systematic errors in the measurement of
helium abundance using only low degree modes. We also look at
models with substantially different abundances, such as stars with Population~II
abundances. We also look at stars beyond the main sequence. The rest of the
paper is organised as follows: we describe the measurement method
in Section~\ref{sec:technique} and give details of the models we
have used in this study in Section~\ref{sec:models}; the results
(including those for solar low degree data) are presented in
Section~\ref{sec:results} and we discuss our conclusions
in Section~\ref{sec:conclusions}.

\section[]{The Technique}
\label{sec:technique}

The amplitude of the oscillatory signal due to the depression in
$\Gamma_1$ in the HeII ionisation zone  is very
small and it may be desirable to amplify it for a proper measurement.
We amplify the signal by taking the second difference of the frequencies,
\begin{equation}
\delta^2\nu_{n,\ell}=\nu_{n+1,\ell}-2\nu_{n,\ell}+\nu_{n-1,\ell}
\label{eq:diff}
\end{equation}
for modes with the same degree. We use modes of degrees $\ell=0$--3.
In addition to amplifying the oscillatory signal, the process of taking
second differences suppresses the dominant smooth trend of the frequencies
to a large extent, making it easier to fit and measure the oscillatory term.
The second differences are then fitted to the form used by \citet{basu97},
but without the degree-dependent terms (which are not relevant
for low degrees), i.e.;
\begin{eqnarray}
&&\delta^{2}\nu=\left(a_1+ a_2\nu +{a_3\over\nu^2}\right) \nonumber\\
&&\qquad\qquad +
  \left(b_1
 +{b_2 \over\nu^2}\right)\sin(4\pi\nu\tau_{\rm {He}}+
 \phi_{\rm {He}})\; \nonumber\\
&&\qquad\qquad +
\left(c_1+{c_2\over\nu^2}\right)
\sin(4\pi\nu\tau_{{\rm CZ}}+\phi_{{\rm CZ}}),
\label{eq:quad}
\end{eqnarray}
where the first term with  coefficients $a_1,a_2,a_3$ defines the smooth part of
the second difference, the second term is the oscillatory signal
from the HeII ionisation zone, and the last term
is the oscillatory signal from the base of the convection zone.
In Eq.~\ref{eq:quad}, the terms $\tau_{\rm He}$ and $\tau_{\rm CZ}$ are the acoustic depths of
the HeII ionisation zone and the base of the convection zone, respectively.
The terms $\phi_{\rm He}$ and $\phi{\rm CZ}$ are the  phases
of the corresponding oscillatory signals.
The parameters $a_1$, $a_2$, $a_3$, $b_1$, $b_2$, $\tau_{\rm {He}}$,
$\phi_{\rm {He}}$, $c_1$, $c_2$, $\tau_{{\rm CZ}}$ and
$\phi_{{\rm CZ}}$ are determined by least-squares fits to the second
differences of the frequencies.

As mentioned earlier, the HeII ionisation zone actually has a small but finite
width and the first derivative of the squared sound speed (after some
scaling) can be better approximated by a triangular function rather than
a discontinuity.
To account for the extra modulation in the oscillatory term 
due to the finite width of the ionisation zone (see Eq.~\ref{eq:factor})
we have also fitted the oscillation with the modulating
factor  $\sin^2(\omega\beta)/(\omega\beta)$ \citep{mt98} in the amplitude
keeping $\beta$ as a free parameter: 
\begin{eqnarray}
&&\delta^{2}\nu=\left(a_1+ {a_2\nu} +{a_3\nu^2}\right) \; \nonumber\\
&&\qquad\qquad +
  \left(b_1\sin^2(2\pi\nu\beta)\over{\nu\beta}\right)
 \sin(4\pi\nu\tau_{\rm {He}}+
 \phi_{\rm {He}})\; \nonumber\\
&&\qquad\qquad +
\left(c_1+{c_2\over\nu}+{c_3\over\nu^2}\right)
\sin(4\pi\nu\tau_{{\rm CZ}}+\phi_{{\rm CZ}}),
\label{eq:mont}
\end{eqnarray}
For most of the cases considered in this paper, $\beta$ is found to 
be of the order of 100~s
and hence, for the range of frequencies included in the fits, the
term $\sin^2(2\pi\nu\beta)$ can be approximated reasonably well
by a quadratic term as assumed in Eq.~(\ref{eq:quad}). 
Thus although the fitting form in Eq.~(\ref{eq:quad}) had been
empirically determined to fit the signature from the
base of the solar convection zone using intermediate degree solar modes,
it is expected to fit the signature from HeII ionisation zone in low
degree modes quite well.

To study the sensitivity
of fitted amplitude and acoustic depth to the fitting form we have
tried one more variation where the oscillatory term due to HeII
ionisation zone in Eq.~(\ref{eq:mont}) is modified to
\begin{equation}
 \left(b_1+{b_2\over\nu}+{b_3\over\nu^2}\right)\sin(4\pi\nu\tau_{\rm {He}}+
 \phi_{\rm {He}})\;. 
\label{eq:quad2}
\end{equation}
This form is obtained by approximating the factor $\sin^2(2\pi\nu\beta)/
(\nu\beta)$ in Eq.~(\ref{eq:mont}) by a quadratic in $1/\nu$. Since
$\beta$ is small such an approximation is reasonably good.
This form differs from Eq.~(\ref{eq:quad}) through the addition of a linear
term in $1/\nu$ in amplitudes.

\begin{figure}
  \begin{center}
    \leavevmode
  \centerline{\epsfig{file=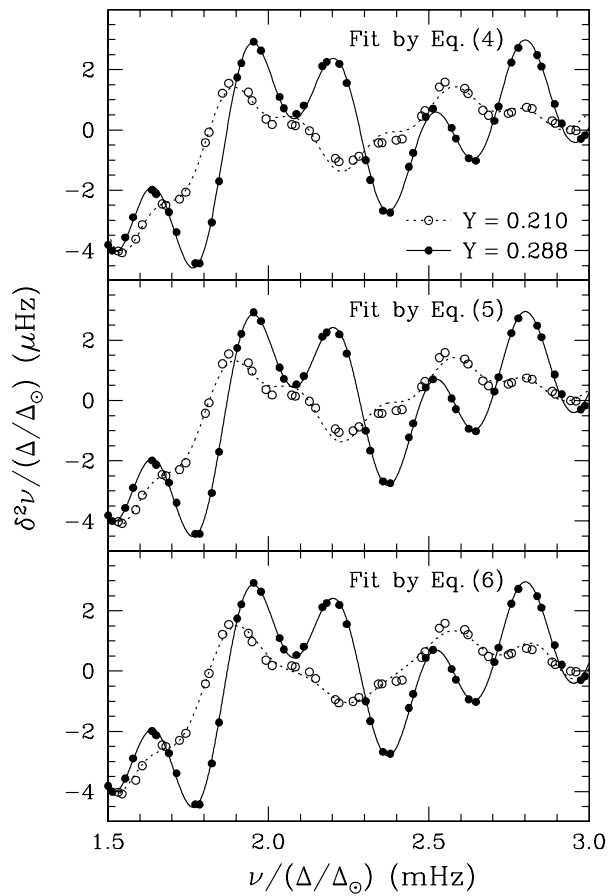,width=8.0cm}}
  \end{center}
  \caption{A sample of the fits to the second differences of the
scaled frequencies. The points are the `data', and the lines the
fits to the points. The examples shown are for a $1.2M_\odot$ model evolved
to a radius of $1.2R_\odot$.}
\label{fig:diff}
\end{figure}

The  fits of the three forms to frequencies generated by
a $1.2 M_\odot$, $1.2R_\odot$ model are shown in Fig.~\ref{fig:diff}.
Each fit is a composite function representing the
smooth trend in the second differences, a dominant slow oscillation due to the
second helium ionisation zone, and a higher frequency oscillation arising from the
base of the convection zone. 
Each panel in the figure shows two fits, one for each of the two extreme
values of the initial helium abundance $Y_0$ that we have chosen for this study.
As can be seen in the figure, the amplitude of the slower oscillation clearly
increases with the amount of helium present in the envelope. We aim to
calibrate this increase in the amplitude of the oscillatory signal from
the HeII ionisation zone against the helium content.
We  do such fits
for each of the stellar models. Since the amplitude of the signal is a function
of frequency, we use the fits to obtain the mean amplitude
over the fitting interval. 
There is not much difference in the fits obtained using the three
forms given by Eqs.~(\ref{eq:quad}--\ref{eq:quad2}), as borne out by 
Fig.~\ref{fig:diff}. This ensures that the amplitudes are not too sensitive to
the particular fitting form chosen.

The frequencies of stellar oscillations roughly scale as
$\sqrt{\bar{\rho}}$, where $\bar{\rho} \propto M/R^3$ is the mean density of 
the star. 
Since the models have different mean densities, it is convenient
to scale the frequencies by the factor
\begin{equation}
f=\sqrt{\bar{\rho}/\bar{\rho}_{\odot}}
\label{eq:scale}
\end{equation}
 before
taking the differences and fitting the signal. However, this requires a knowledge
of $M/R^3$ for the stars to be studied, which may not be known for
all stars that may be targets of asteroseismic study.
This factor can actually be
estimated from the seismic data using the large frequency separation,
\begin{equation}
\Delta_{n,\ell}=\nu_{n+1,\ell}-\nu_{n,\ell} 
\label{eq:delta}
\end{equation}
\citep[see e.g.,][]{ma01,jcd03}.
We  find that the averaged large frequency separation   scaled by the
factor in Eq.~(\ref{eq:scale}) is around 135~$\mu$Hz for all the main-sequence stellar 
models in the
range of masses considered in this study.
We therefore, scale the frequencies using the average frequency
separation to avoid the use of the potentially unknown factor $M/R^3$.
We have chosen a frequency range of 1.5~mHz to 3.0~mHz after
scaling by $\Delta/\Delta_\odot$, where $\Delta$ is the mean
large frequency separation and $\Delta_\odot=135\,\mu$Hz
is the value for the Sun. 
The choice of the frequency range is dictated by practical reasons. 
At higher frequencies the oscillatory signal
due to the HeII ionisation zone becomes small. Furthermore,
the observational error is expected to increase in higher frequency modes,
as is the case for solar data. 
At lower frequencies there is significant departure from the fitting forms
assumed in this work since the low $n$ modes can not be  approximated well
by the asymptotic formulae used in obtaining the oscillatory form \citep{mon96}.

It can be  shown that the process of taking
the second differences magnifies the amplitude of the oscillatory
functions by a factor 
\begin{equation}
4\sin^2(2\pi \tau\Delta)
\label{eq:mag}
\end{equation}
where $\Delta$ is the mean
large frequency separation. In order to obtain the real 
amplitude of the oscillatory signal, i.e., the amplitude
of the oscillations in the frequencies, we divide the amplitude obtained for
the second differences by this factor.
This conversion back to frequencies is found to remove some variation 
in the amplitude between different models \citep{ma01}.
All the amplitudes presented in  this work are the amplitudes
of the oscillatory function in the frequencies that we obtain by
converting the fitted amplitudes in the second differences
using the conversion factor in Eq.~(\ref{eq:mag}).
We find that for the  HeII ionisation
zone, this factor is not very different from unity and hence the amplitudes
are similar for both frequencies and the second differences of
frequencies. This magnification is however quite large
for the signal from the 
base of the convection zone, making it easier to
fit the two oscillatory terms simultaneously in the second differences.

By fitting the oscillatory term in the second differences,
it is  possible
to determine the characteristics of the HeII ionisation zone, like its
acoustic depth and the thickness and amplitude of the depression in $\Gamma_1$ 
from the fitted parameters. One would
expect the amplitude of the oscillatory signature to  increase with 
helium abundance and hence it can
be calibrated using stellar models with known $Y$. \citet{mig03}
have suggested that the area of the bump in $\Gamma_1$ should be
a measure of helium abundance. The product of amplitude with
half-width $\beta$  provides a measure of this area. We find that
it is difficult to estimate $\beta$ reliably by fitting the
oscillatory term.
We find a large scatter in the results if the product of the
amplitude with $\beta$ is used, which makes it difficult to determine
$Y$ and hence we do not use it in our study.
Instead, we try to find a relation directly between the amplitude of
oscillatory term and the helium abundance.

\section[]{Stellar Models}
\label{sec:models}

We use the technique described in the previous section
to study the properties of the oscillatory signal in the frequencies
of low degree modes and its relationship with helium abundance
in the stellar envelope using a large number of stellar models.
We restrict ourselves to stars with masses close to the
mass of the Sun, namely, the range 0.8--1.4~$M_\odot$.
The upper limit of the mass range
is set by the fact that it is difficult to extract the oscillatory
signal in more massive stars. This happens  because the acoustic depths
of  the base of the convection
zone and that of the HeII ionisation zone  are very similar  in higher mass
stars,
and the two oscillatory signals, one from the convection zone base and the
other from the helium ionisation zone, interfere with each other. 
The lower mass limit is dictated by the fact that the  amplitude of the oscillatory signal 
reduces with mass, though there is no particular difficulty in
determining the signal as we are using exact model frequencies. 
In actual practice the lower limit on the stellar
mass that can be studied will be determined by
the precision to which the frequencies can be
determined from the future asteroseismic missions.
Since observed frequencies will have uncertainties associated with them,
we have  examined the effect of   frequency errors  on the
fits in Section~\ref{sec:results}.
We study stellar models evolved to different ages on the main
sequence for masses $0.8M_\odot$, $1M_\odot$, $1.2M_\odot$ and
$1.4M_\odot$. We have also constructed a few stellar models which
have evolved beyond the main sequence and are on the subgiant branch.

The stellar models were constructed using YREC, the Yale Rotating
Evolution Code in its non-rotating configuration \citep{gue92}.
These models use the OPAL  equation of state \citep{rn02},
OPAL opacities \citep{ir96}, low temperature
opacities of \citet{af94} and nuclear reaction rates
as used by \citet{ba92}.
The models take into account diffusion of helium and heavy
elements, using the prescription of \citet{tbl94}.

\begin{figure*}
  \begin{center}
    \leavevmode
  \centerline{\epsfig{file=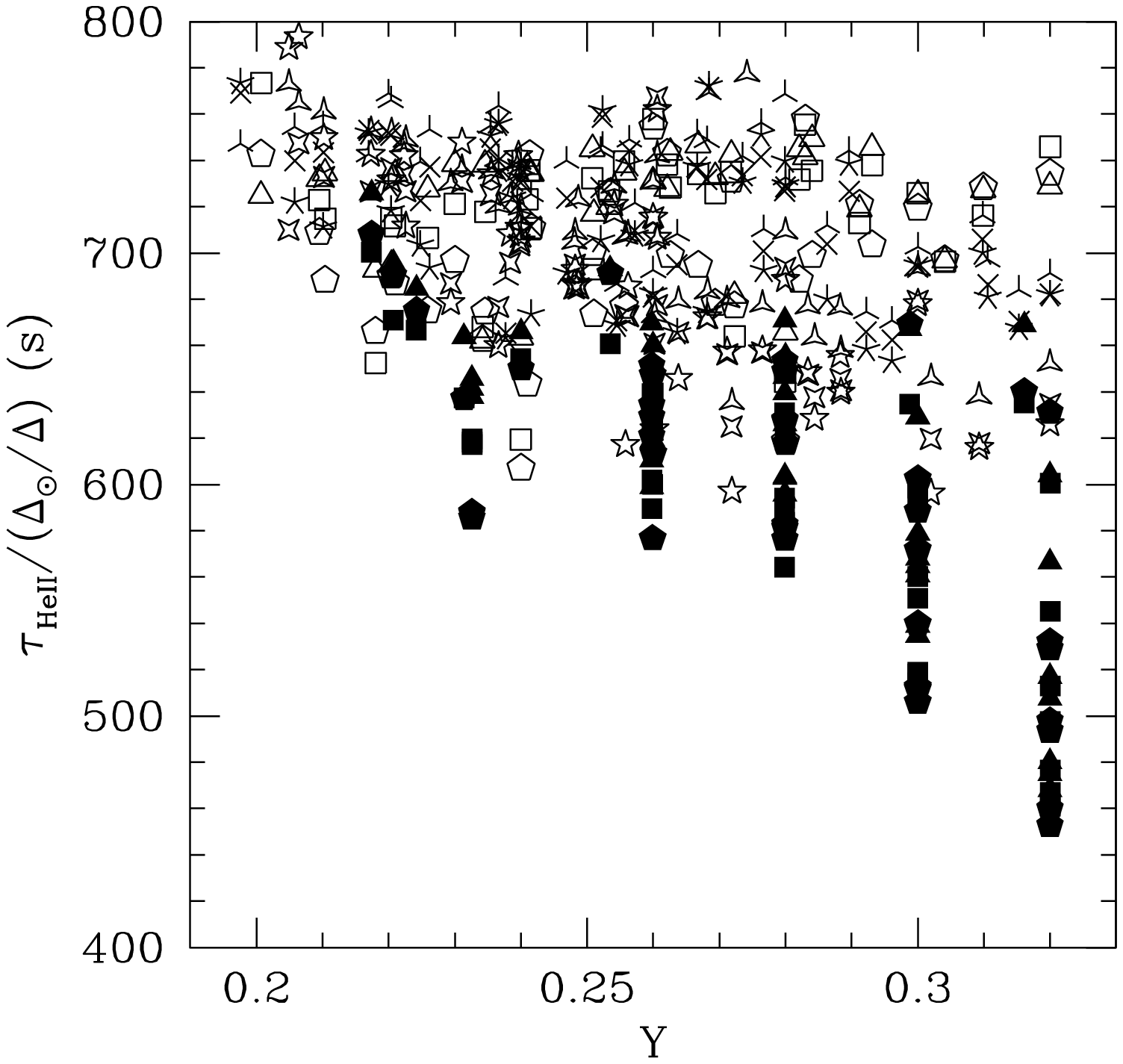,width=8.0cm}\hspace{1 cm}
  \epsfig{file=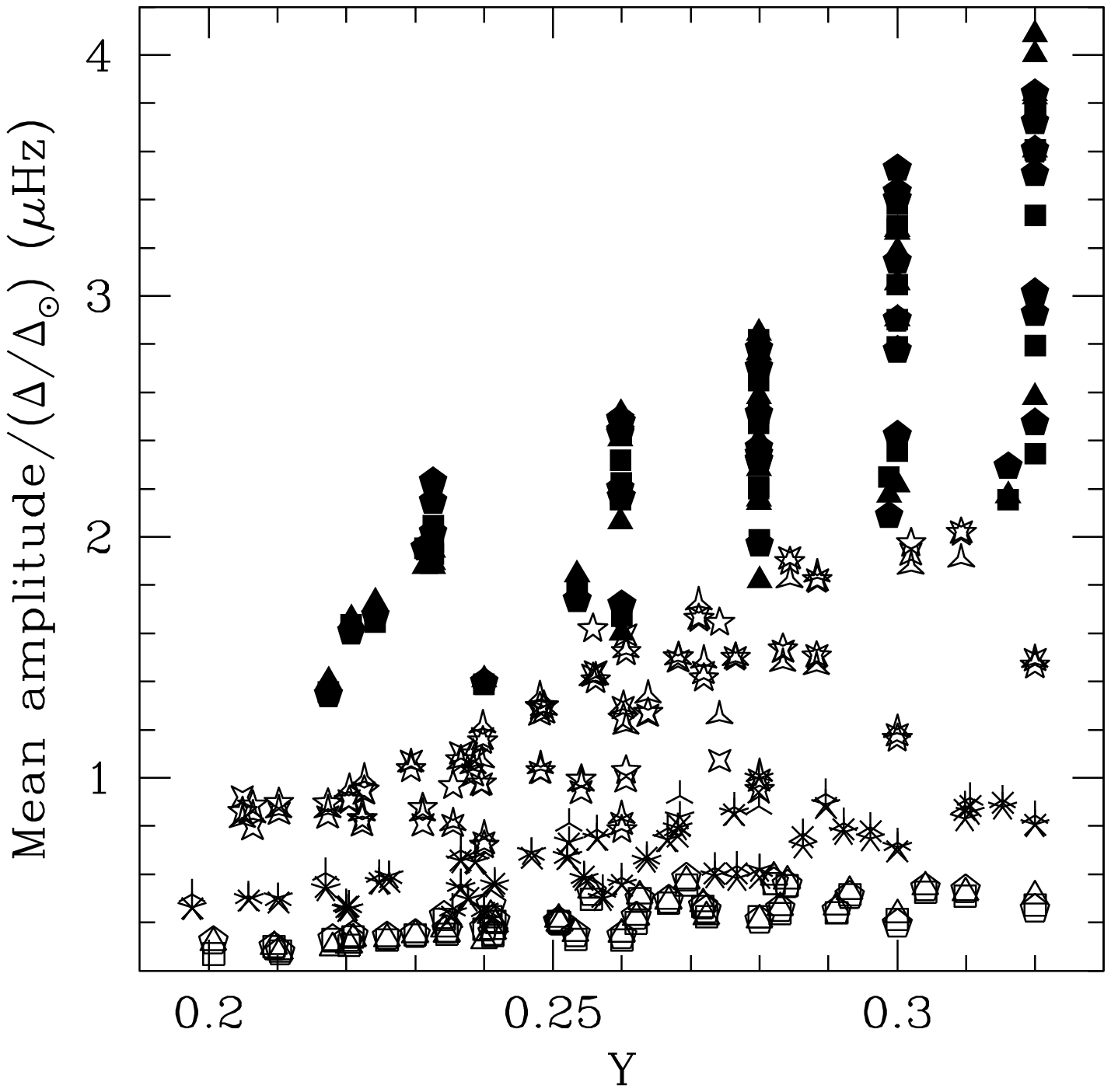,width=8.0cm}}
  \end{center}
  \caption{The scaled acoustic depth $\tau_{\rm He}$ and the mean amplitude
of the oscillatory part in frequency due to the HeII ionisation zone
in the scaled frequency range of 1.5--3.0~mHz
are shown as a function of the helium abundance in the envelope.
The open symbols represent $0.8M_\odot$ models, skeletal symbols show
$1M_\odot$, starred for $1.2M_\odot$ and filled symbols show $1.4M_\odot$
models. For each model the results for the three independent fits using
Eqs.~(\ref{eq:quad}, \ref{eq:mont}, \ref{eq:quad2}) are
shown by 3, 4 and 5 sided symbols respectively.}
\label{fig:tau}
\end{figure*}

Most of the models have been constructed with Population~I abundances.
These models have an initial heavy element abundance $Z_0=0.022$ and were constructed
with a mixing length parameter $\alpha=2.1$. The initial heavy element abundance and the
 mixing length
are the same as those used to construct calibrated  standard solar models using YREC 
\citep{win02}. The value of the initial helium abundance
$Y_0$ is varied: we use 5 values of $Y_0$ for the Population~I models,
$Y_0=0.24$, 0.26, 0.28, 0.30 and 0.32.  
Diffusion and gravitational settling of helium causes the helium abundance
in the stellar envelope to change with time. Therefore, the helium
content in the envelope of a star still evolving on the main sequence 
will decrease progressively with age from its initial  (or
zero-age main sequence) helium abundance $Y_0$. Beyond the main sequence
the dredge-up of helium begins to increase the helium abundance in the
envelope.
Thus the amplitude of the HeII signal will
not reflect the initial abundance $Y_0$ of the models; it will depend on the
abundance of helium in the stellar envelope at the stage we examine it.
As a result we always use the actual, rather than the initial, helium abundance 
in the envelope to compare and
calibrate the amplitude of oscillatory signal.
To investigate systematic
errors caused by model parameters, we have also constructed models with a slightly
different value of the initial heavy element abundance ($Z_0=0.018$) and some models
with a slightly different value of the mixing length parameter ($\alpha=1.7$).

To test whether the method outlined in this paper also applies to stars of very
different heavy element abundances, we have two other sets of models. One set with
initial abundance $Z_0$ of 0.007 (the abundance of stars like $\tau$~Ceti) and another set
with Population~II abundance of $Z_0=0.001$. Models for these two sets
were constructed with initial $Y$ of 0.22, 0.24, 0.26, 0.28 and 0.30.
For each of the stellar models we calculate the frequencies of
low degree modes and attempt to extract the oscillatory term as
explained in the previous section.

From our experience of working with solar data, we expect that one
of the important sources of uncertainty is the equation of state (EOS)
which determines the width and the depth of the dip in $\Gamma_1$.
The role of
the equation of state has been extensively studied for the solar case
\citep{ab94,rich98,basu98} by looking
at models constructed with  OPAL, the Mihalas, Hummer \& D\"appen (henceforth 
MHD) EOS \citep{hm88,mdh88,dap88}
as well as other simpler EOS.
\citet{mt98} have studied the influence of equation of state
in stellar case too and find that the half-width $\beta$ of the HeII
ionisation zone is determined by the EOS.
Helioseismic studies
have shown that in the helium ionisation zone, the OPAL EOS
is a better representation of stellar EOS \citep[see e.g.,][]{ba95,bc97,bdn99}
and hence, in this work we restrict ourselves to models
with OPAL EOS, though we do study the effect of the EOS using a few
$1M_\odot$ models constructed with the MHD equation of state. This should
give a reasonable estimate of the uncertainty associated with the EOS.

\section[]{Results}
\label{sec:results}

\subsection[]{Population~I Main Sequence Models}
\label{subsec:std}

\begin{figure*}
  \begin{center}
    \leavevmode
  \centerline{\epsfig{file=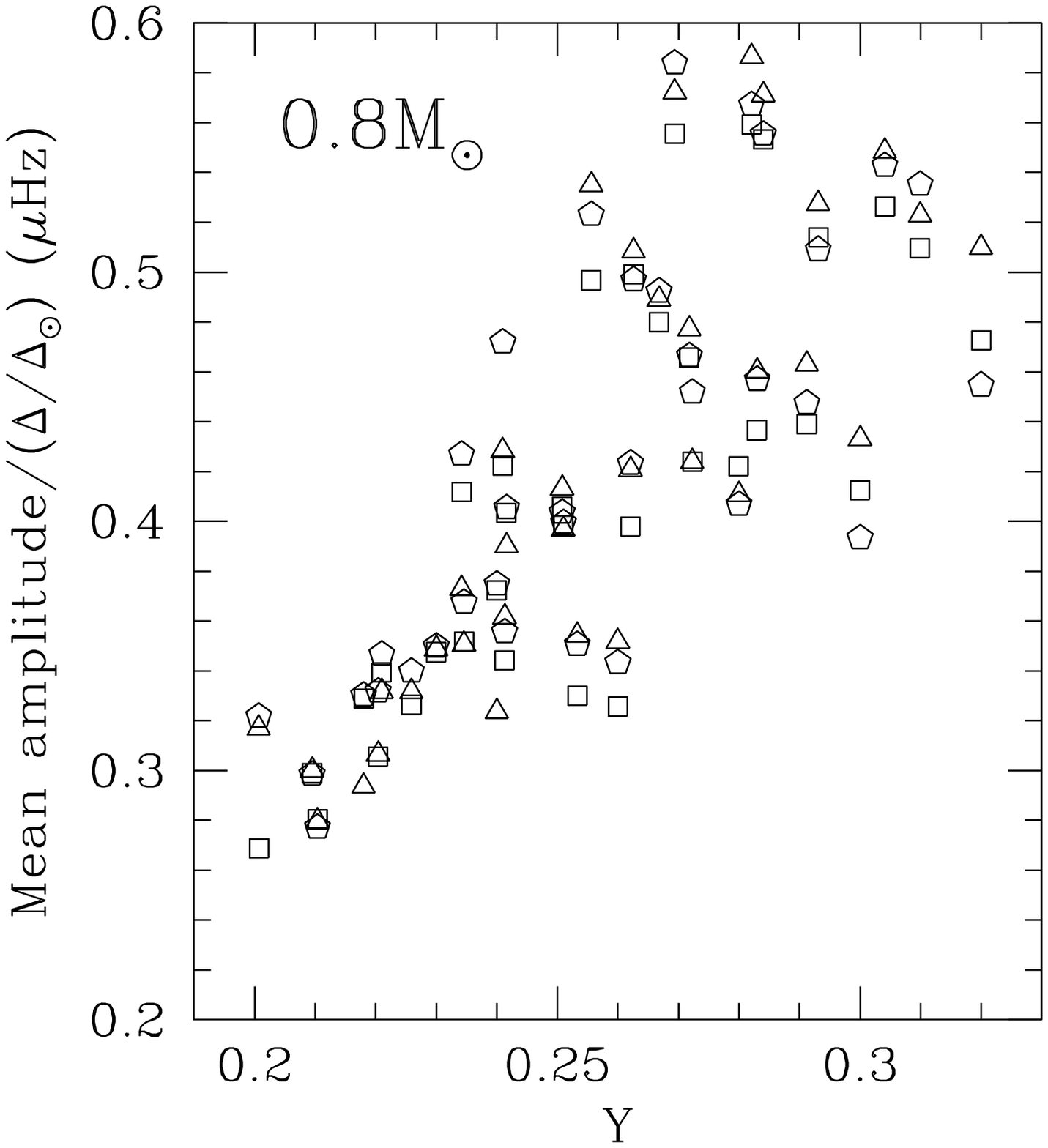,height=4.4cm}\hspace{0.4 cm}
  \epsfig{file=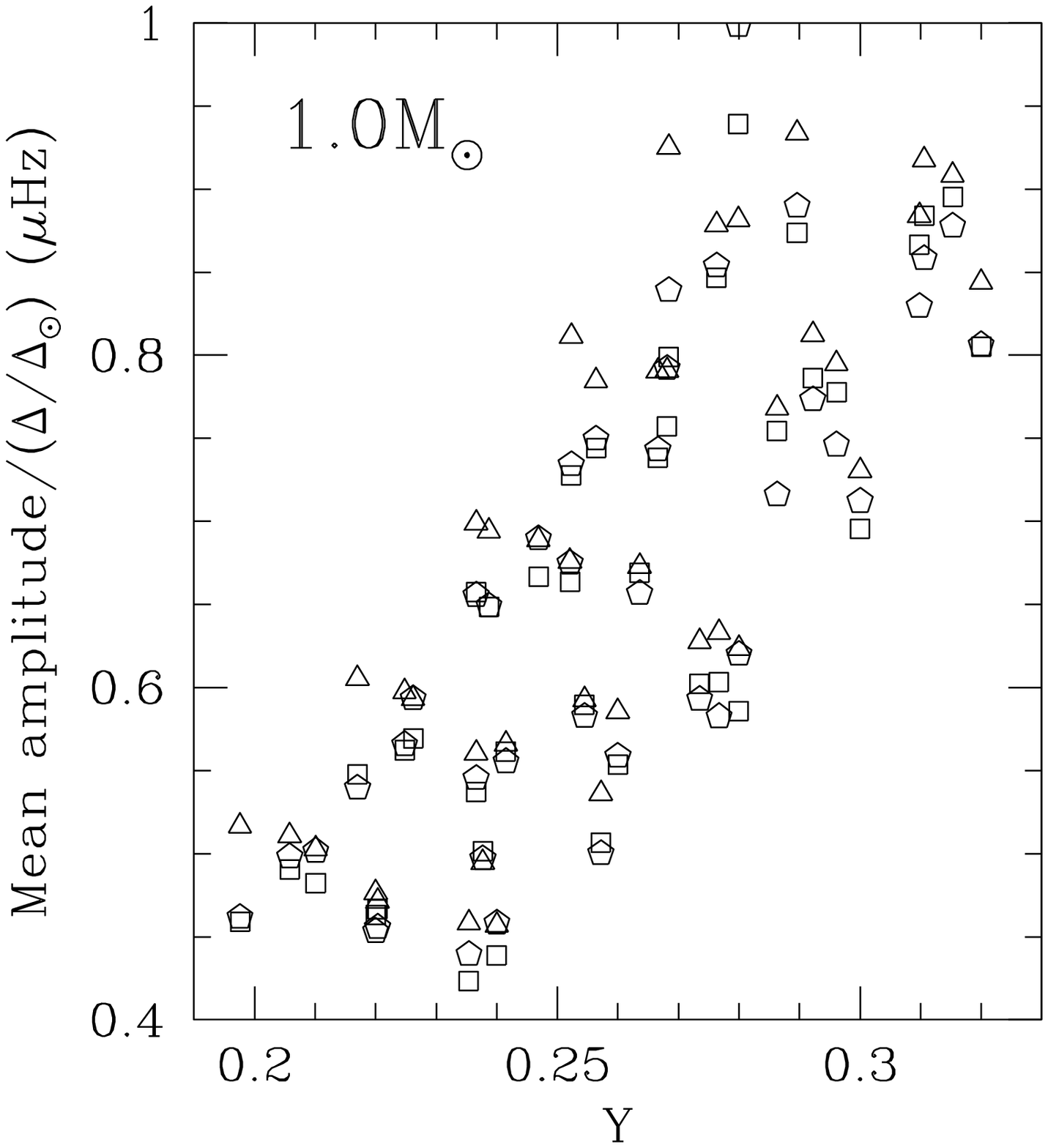,height=4.4cm}\hspace{0.4 cm}
  \epsfig{file=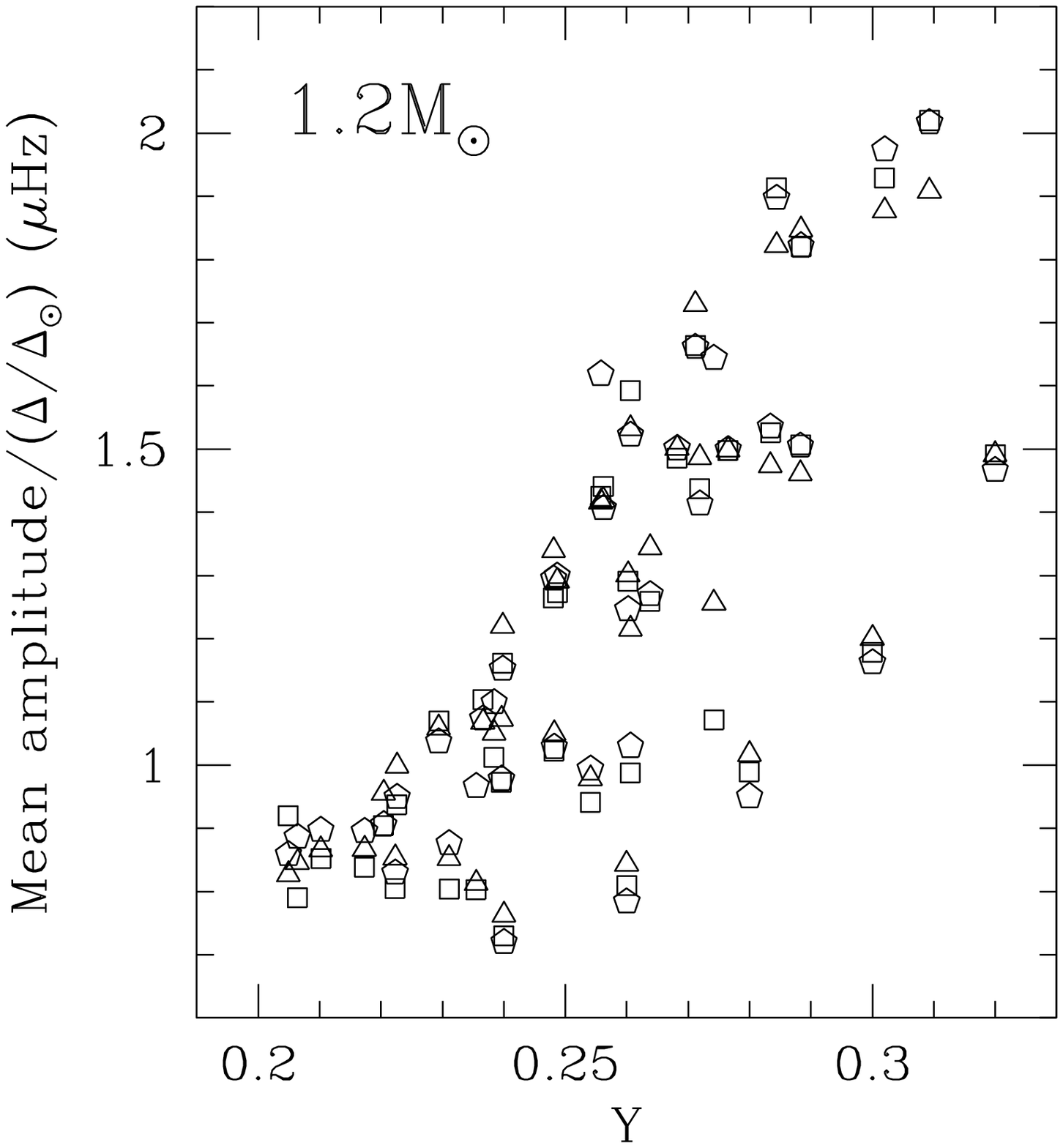,height=4.4cm}\hspace{0.4 cm}
  \epsfig{file=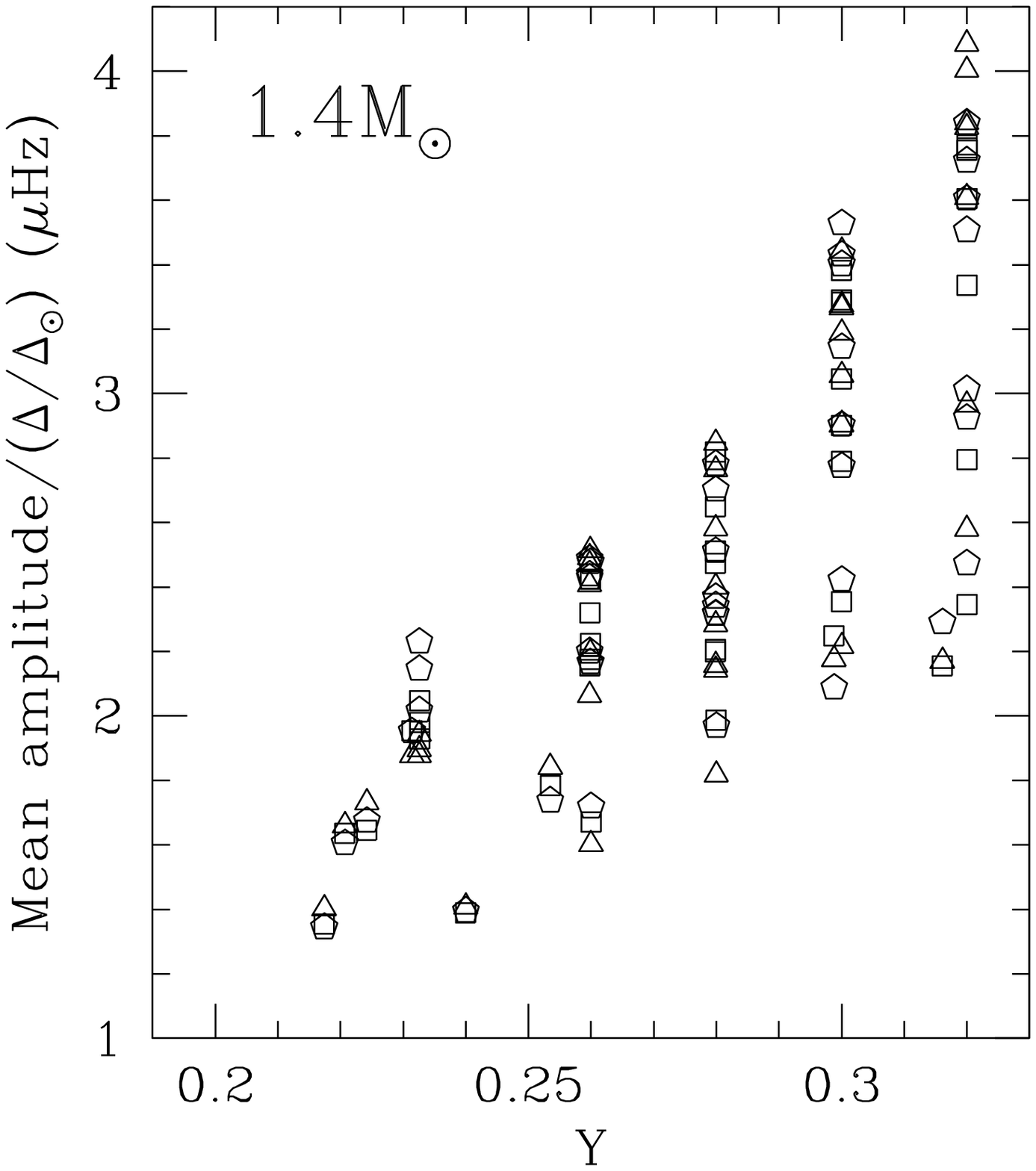,height=4.4cm}}
  \end{center}
  \caption{The mean amplitude
of the oscillatory part in frequency due to the HeII ionisation zone
is shown as a function of the helium abundance in the envelope.
Each panel shows the result for models with a fixed mass.
The triangles, squares and pentagons show the results using fitting
functions (\ref{eq:quad}), (\ref{eq:mont}) and (\ref{eq:quad2})
respectively.}
\label{fig:ampmass}
\end{figure*}

Fig.~\ref{fig:tau}  shows the acoustic depth
and the average amplitude of the oscillatory signal due to the HeII ionisation zone
for the standard set of  main sequence models used in the study.
It can be seen
that the acoustic depth $\tau_{\rm He}$ is within a factor of two
for all models, while the amplitude varies over a large
range. It should be noted that all values of the acoustic depth and
the amplitude listed in this paper are those obtained
after scaling by the frequency separation.
The actual values of $\tau_{\rm He}$ may show significant variation
due to variation in $M/R^3$ or equivalently in $\Delta$.
On average, the amplitude appears to increase with
$Y$, the helium abundance in the stellar envelope, but  for a given 
helium abundance, the amplitude also increases with stellar mass.
This figure combines the results obtained using different fitting
forms and it can be seen that there is a good agreement between
different results, though for $1.4M_\odot$ models the difference
tends to become large as the fits are not very good for many of these
models.
The mean difference in amplitudes obtained using two different fitting
forms is found to be about 0.02 $\mu$Hz for stellar models with
masses $0.8M_\odot$, while for $1M_\odot$ and $1.2M_\odot$ models
it is about 0.06 $\mu$Hz. As we shall see later, these differences are
smaller than the
errors that may be expected to arise from uncertainties in observed
frequencies. For $1.4M_\odot$ models the mean difference in amplitudes
between two fitting forms is about 0.18 $\mu$Hz, which is comparable
to the errors expected to be caused by uncertainties in the observed frequencies. 
To show the variation of amplitude with $Y$ more clearly,
Fig.~\ref{fig:ampmass} shows the
amplitude as a function of $Y$ for each mass separately.
There is considerable scatter around the mean
trend due to factors like age, and hence the evolutionary state.

\begin{figure*}
  \begin{center}
    \leavevmode
  \centerline{\epsfig{file=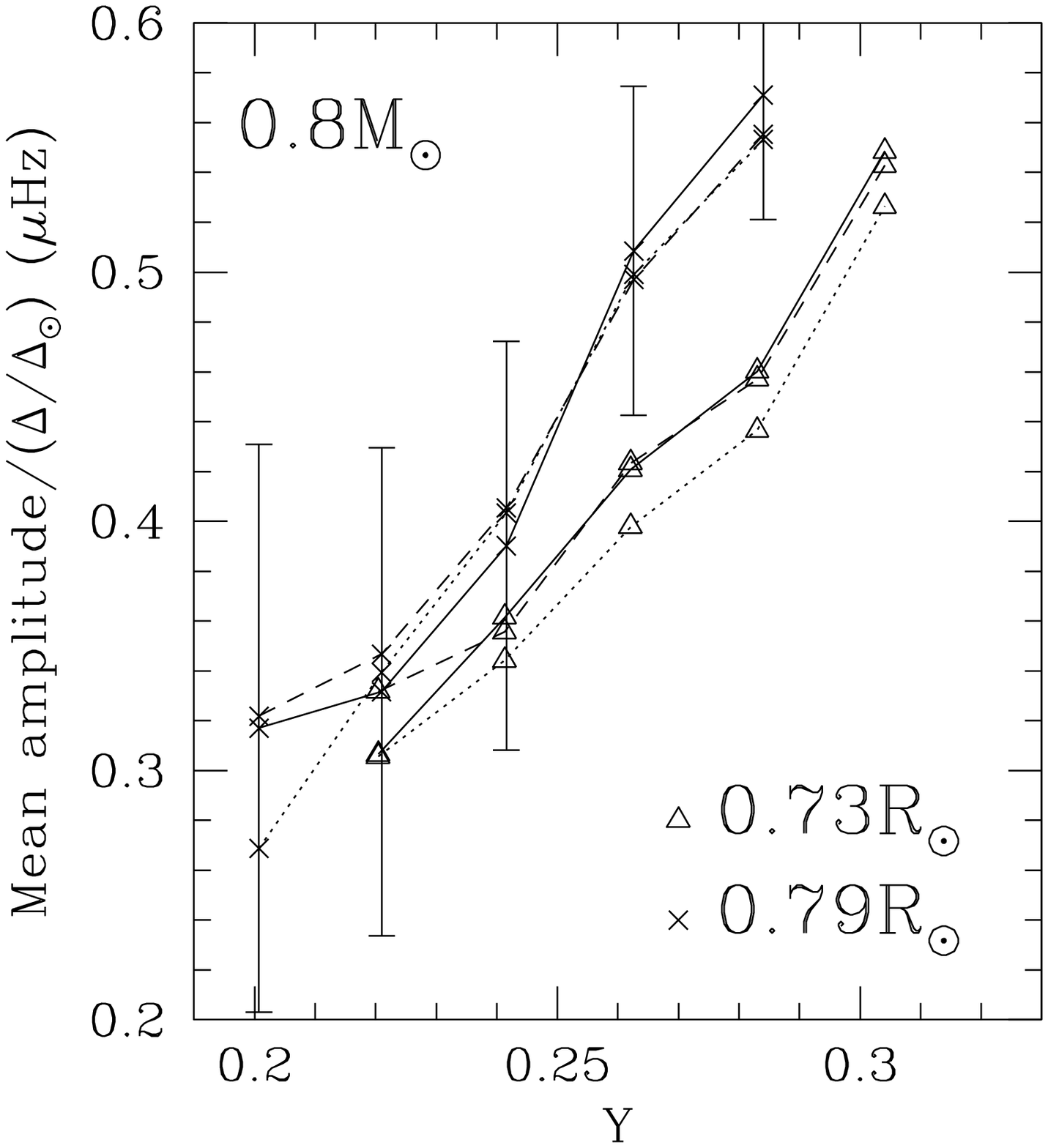,height=4.4cm}\hspace{0.4 cm}
  \epsfig{file=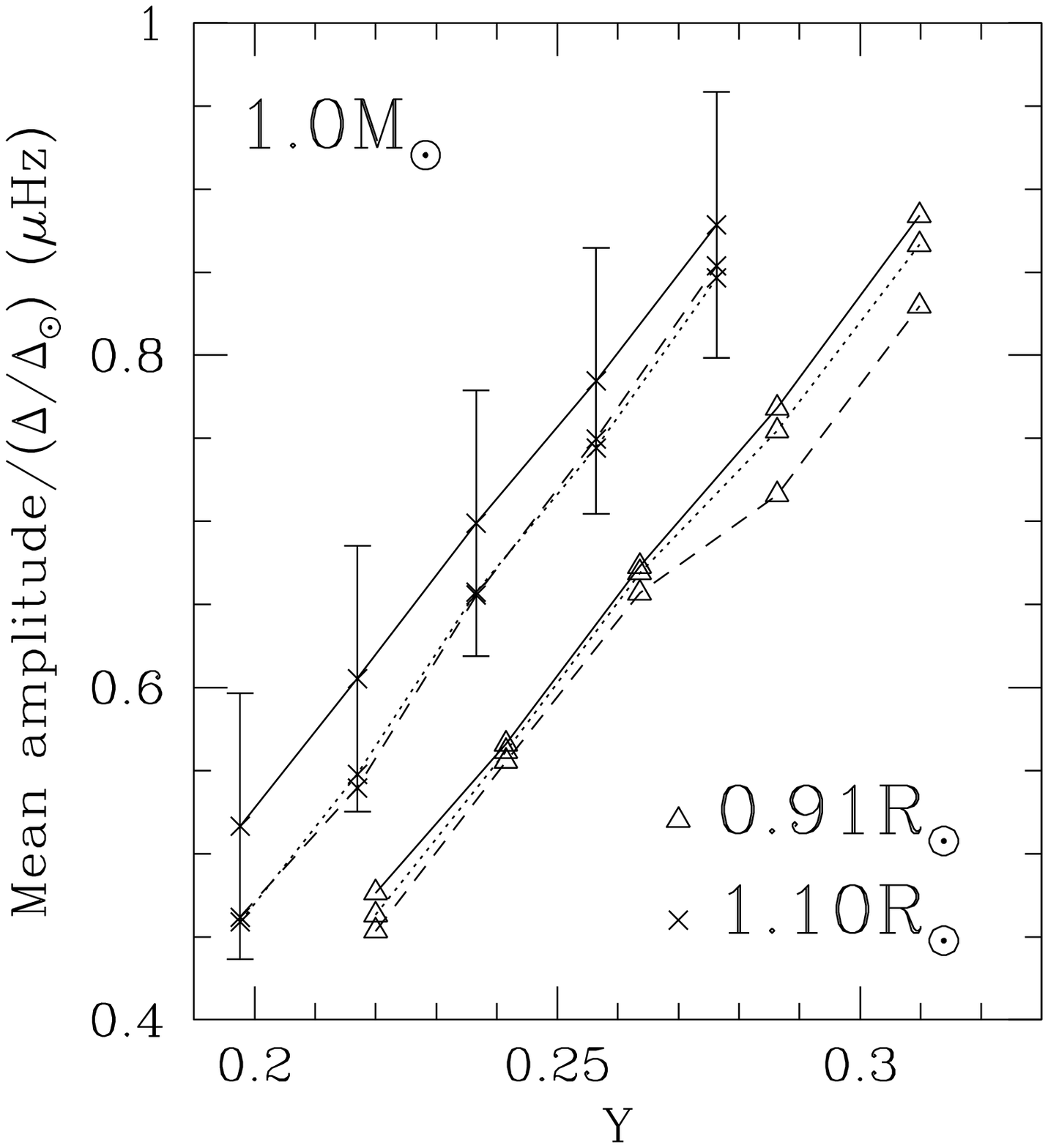,height=4.4cm}\hspace{0.4 cm}
  \epsfig{file=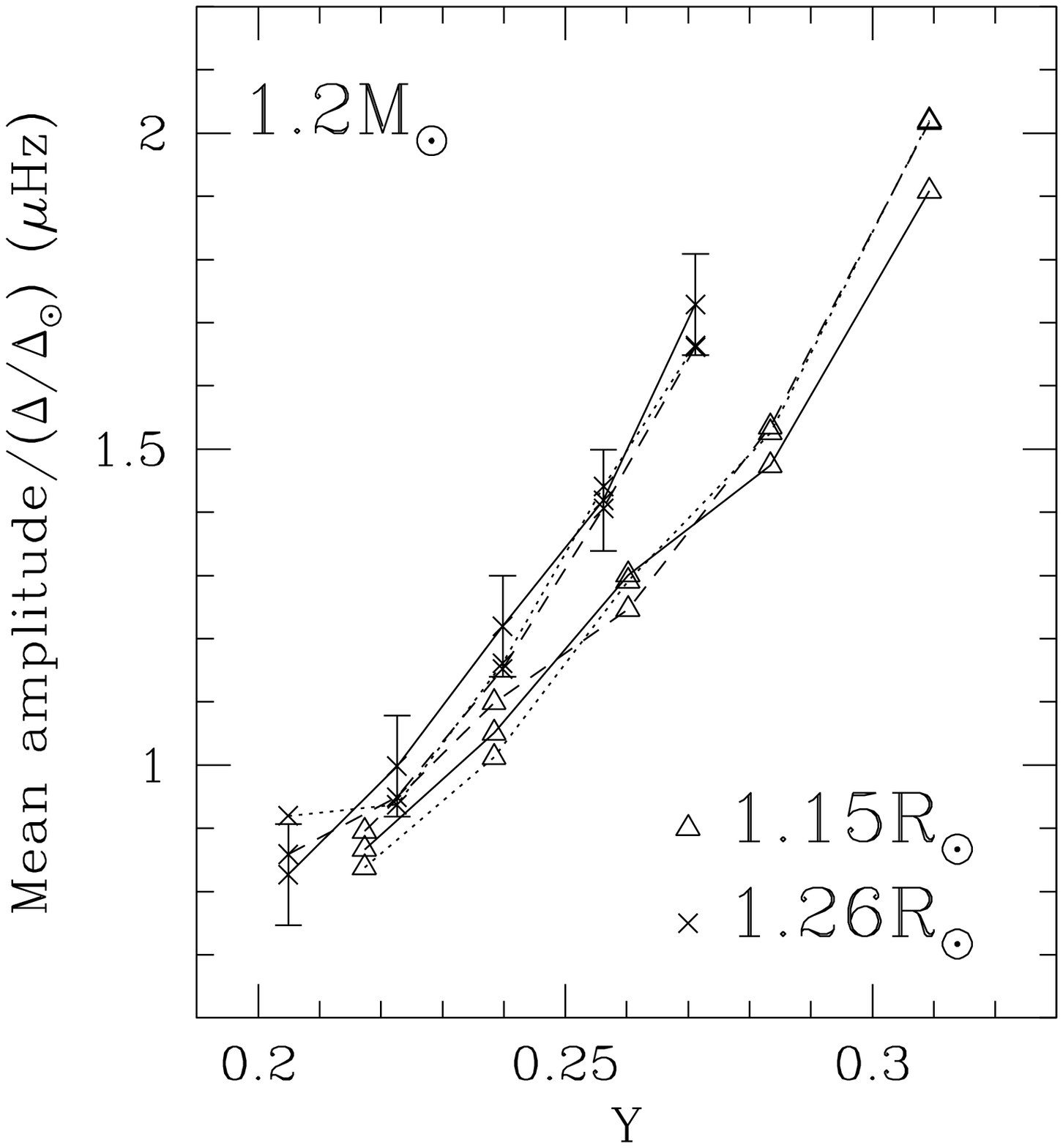,height=4.4cm}\hspace{0.4 cm}
  \epsfig{file=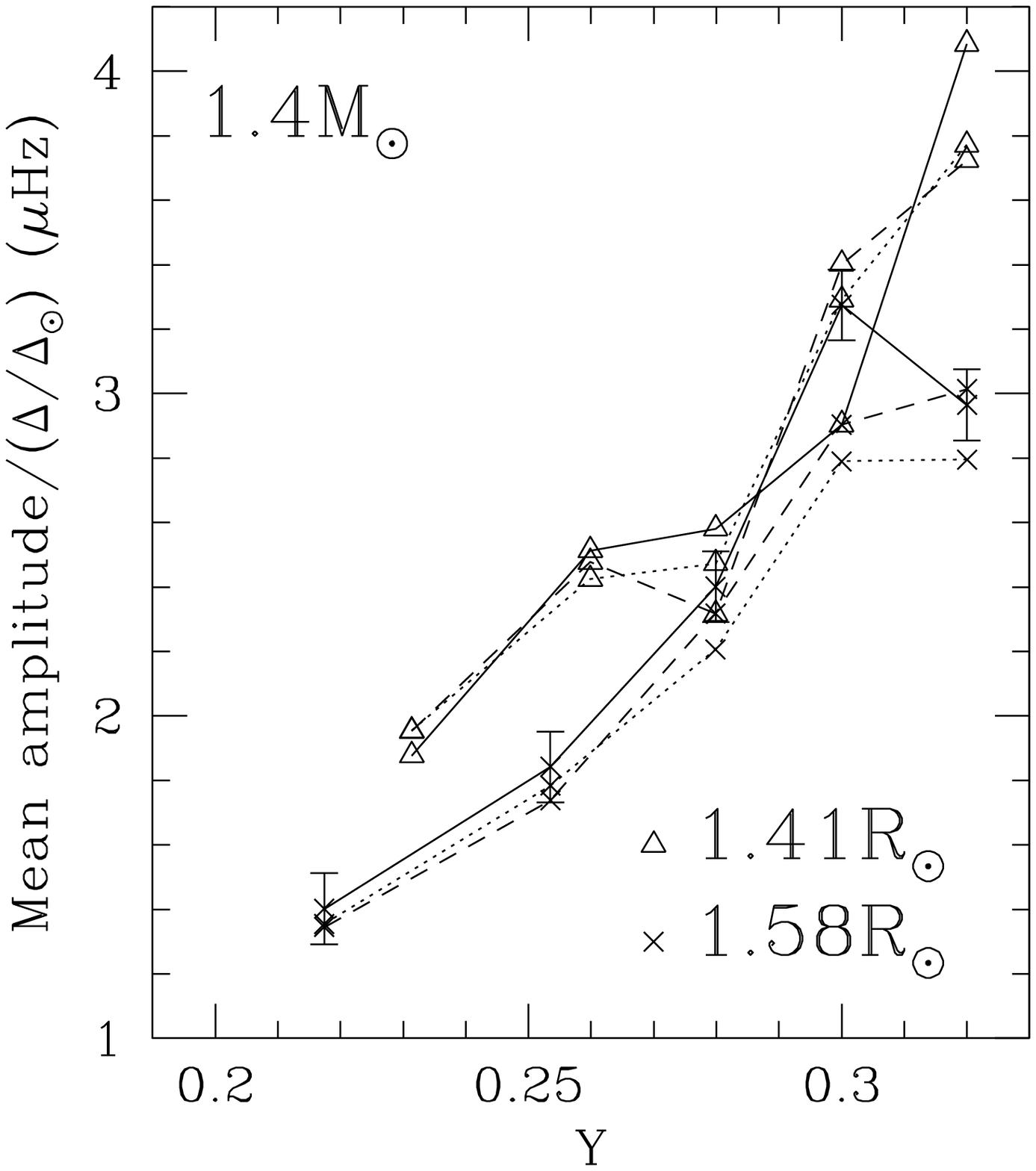,height=4.4cm}}
  \end{center}
  \caption{The mean amplitude
of the oscillatory part in frequency due to the HeII ionisation zone
is shown as a function of the helium abundance in the envelope for
models with fixed mass and radius. The different masses and radii are
marked in each panel. The solid, dotted and dashed lines
show the results for fits using Eqs.~(\ref{eq:quad}), (\ref{eq:mont})
and (\ref{eq:quad2}) respectively. For sake of clarity, estimated errors
are shown on only one set of points.}
\label{fig:radius}
\end{figure*}

The dependence of the amplitude on factors other than the helium
abundance complicates the determination of the helium abundance of
a given star.
Our task is made easier if conventional observations combined with
modelling give us a reasonable idea of the  radius, $R$  and mass $M$
of the star. 
If only the radius is known,  we can estimate
$M/R^3$ from the frequency separation and combining that with the known
radius will give an estimate of mass. Similarly, if only the stellar
mass is known independently, we can use the frequency separation to
estimate the stellar radius.
This would enable us to make calibration models with a given radius and mass.
A plot of the mean amplitude as a function of $Y$ for models with
given masses and
radii is shown in Fig.~\ref{fig:radius}. In this case all we would need to do
to determine $Y$ 
is to determine the mean amplitude from the observations and
interpolate. Models of the same mass but different initial helium
abundances will have different ages when they have the same radius, as we
shall see in section~\ref{subsec:sun}, this does not affect the calibration
process.

Fig.~\ref{fig:radius} also shows an estimate of the uncertainties in $Y$
expected as a result of observational errors. Shown in the figure are the error bars
on a few representative points, which are
obtained assuming a constant relative error of 1~part in $10^4$ in the frequencies.
This level of error is not unrealistic given that even ground-based
observations have almost reached this level of precision \citep[see][]{bc02}.
These error bars have been calculated using a Monte-Carlo simulation
where random errors with specified standard deviation were added to
the exact frequencies before fitting the data. The results of the fits
to the error-added frequencies were then used to estimate
the variance in each fitted parameter to calculate the expected error.
The precision to which we can use the amplitude to determine $Y$ increases
with increase in mass and age. From the error bars it appears that the
expected error in $Y$ would range from less than 0.01 for $1.2M_\odot$
models to about 0.03 at $0.8M_\odot$.  To get better precision
for $0.8M_\odot$  stars we will need to measure the frequencies to a
relative accuracy of better than $10^{-4}$.
The error in the scaled acoustic depth $\tau$
of the HeII ionisation zone is about 20--40~s except for $0.8M_\odot$
models where it is about 100~s.
The above results were obtained assuming we had modes with degrees
of 0,1,2 and 3. The uncertainty in the $Y$ determination increases
if $\ell=3$ modes are unavailable, though the increase is very marginal
and will not affect any of the results. Obviously, there would be
additional errors due to uncertainties in our knowledge of stellar
mass and radius. This error can be estimated by considering all
models within the permitted range of mass and radius.

It should be noted that the points  shown
in figures have been obtained using the exact model frequencies,
only the extent of the error bars were obtained
from the Monte-Carlo simulations.  From simulations it is found that when random errors
are added to the frequencies, the estimated amplitudes increase 
by an amount comparable to the estimated errors. This effect has been
seen earlier for helioseismic data \citep[see e.g.,][]{basu97}.
Therefore for the purposes of calibration once real data are available,
the calibration curve should be obtained after adding the random errors
to the frequencies of the calibration models.
We have not done that in this work as in that case the figures will
depend somewhat on the estimated errors in frequencies.

\begin{figure}
  \begin{center}
    \leavevmode
  \centerline{\epsfig{file=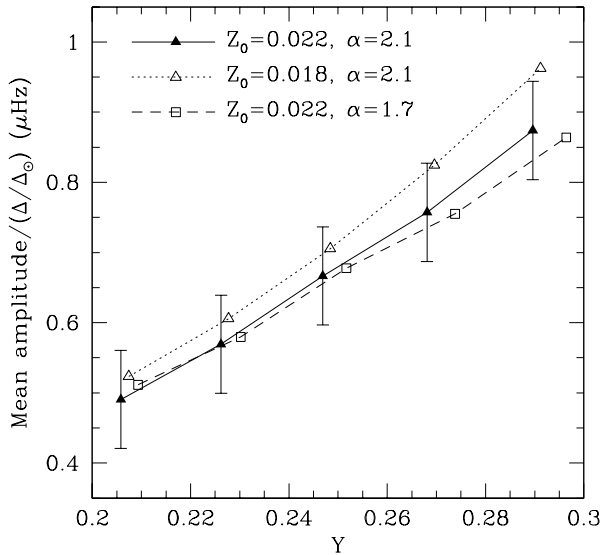,width=8.0cm}}
  \end{center}
  \caption{The mean amplitude
of the oscillatory part in frequency due to the HeII ionisation zone
is shown as a function of the helium abundance in the envelope for
$1M_\odot$ stars that have a  radius of $1R_\odot$.
The results
obtained for models with a slightly different initial  heavy element abundance $Z_0$ and mixing length
parameter $\alpha$ are compared.
Representative errors in
the amplitudes are indicated on one set of points.
All results were obtained by using the function in Eq.~(\ref{eq:mont})
to fit the data.}
\label{fig:zalp}
\end{figure}

To study the influence of the  initial  heavy element
abundance, $Z_0$, and of the mixing length parameter, $\alpha$, on the
amplitude of oscillatory signal we have constructed stellar models
with different values $Z_0$ and $\alpha$ for stars with the same masses and
radii shown in Fig.~\ref{fig:radius}.  Figure~\ref{fig:zalp} shows the results for 
$1M_\odot$ stars that have the same radius ($1R_\odot$) but have
different $Z_0$ and $\alpha$. It can be seen that the variation
in amplitude due to reasonable uncertainties in $Z_0$ and $\alpha$ are
smaller than the variation due to $Y$, and also smaller than the
error estimates for a relative error of $10^{-4}$ in frequencies.

\begin{figure}
  \begin{center}
    \leavevmode
  \centerline{\epsfig{file=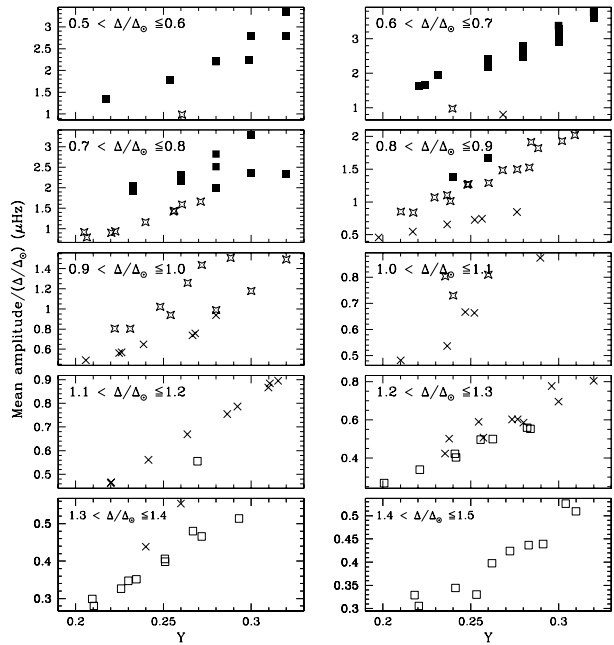,width=8.0cm}}
  \end{center}
  \caption{The mean amplitude
of the oscillatory part in frequency due to the HeII ionisation zone
is shown as a function of the helium abundance in the envelope.
Each panel shows the results for a given bin in the ratio of the large frequency
separation $\Delta/\Delta_\odot$. For clarity only results obtained using
Eq.~(\ref{eq:mont}) are shown. Open squares, crosses, stars and
filled squares displays stellar models with masses $0.8M_\odot$,
$1M_\odot$, $1.2M_\odot$ and $1.4M_\odot$ respectively.}
\label{fig:del}
\end{figure}

In the above discussion we have assumed that either the mass or the radius
of the star is known independently. This is a reasonable assumption since
most of the targets of asteroseismic missions are likely to be nearby stars
for which at least, one of these quantities is known. However, if for
some star neither the mass nor the radius is known, then from the frequency
separation we can only estimate the ratio $M/R^3$. To examine the
implication in this case, we show in Fig.~\ref{fig:del}  
 the amplitude as a function of
$Y$ for all models in a given bin of the mean large frequency separation
$\Delta$. It can be seen that there is some scatter due to variation
in mass and radius, but in most cases the trend of increasing amplitude
with $Y$ is very clear and can be calibrated to calculate the helium
abundance. Of course, the systematic errors will be larger in this
case.

\subsection[]{Results for the Sun}
\label{subsec:sun}

\begin{figure}
  \begin{center}
    \leavevmode
  \centerline{\epsfig{file=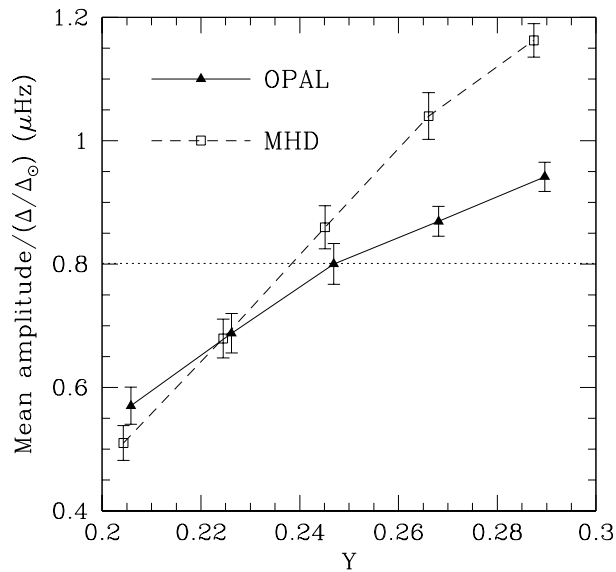,width=8.0cm}}
  \end{center}
  \caption{The mean amplitude
of the oscillatory part in frequency due to the HeII ionisation zone
is shown as a function of the helium abundance in the envelope for
models with $M=1 M_\odot$ and $R= 1R_\odot$. The horizontal line shows
the results obtained using observed frequencies from MDI
instrument for the Sun. The two lines show the results using OPAL
and MHD EOS as marked in the figure. These results are obtained
using the function in Eq.~(\ref{eq:quad}) to fit the data.}
\label{fig:sun}
\end{figure}

In order to test this technique we can try to estimate the helium
abundance in the solar envelope using only the low degree modes. 
We use the frequencies 
obtained from the first 360 days of observations  by the MDI instrument on
board the SOHO spacecraft \citep{sch98}. The observed frequencies with degrees
of $\ell\le3$ and frequencies in the range  $1.5\le\nu\le3.0$~mHz were fitted
to obtain the mean amplitude of the oscillatory signal due to the HeII
ionisation zone, and the result is  shown by the dotted horizontal line in
Fig.~\ref{fig:sun}. To infer the helium abundance in the Sun
we have used $1M_\odot$, $1R_\odot$ models for the calibration.
It should be noted that while these models have the same mass and
radius as the Sun, these calibration models are not solar models, the
models have different ages and luminosities. To the frequencies of each
model we  added fifty realisations of the observed errors,  and for each realisation
we determined the amplitude of the oscillations. We have plotted  the mean amplitude 
for each model as a function of the envelope helium abundance of the model
in Fig.~\ref{fig:sun}. All these fits use the same
set of modes that were used to extract the oscillatory signal from
the observed data.  To study the sensitivity of results to uncertainties
in the EOS, the figure also shows the results of a set of 
models constructed with the MHD equation of state. The result of the
observations is shown as the horizontal line and we can see that
we obtain a helium abundance of $0.239\pm0.005$ with the MHD models and
$0.246\pm0.006$ with the OPAL models. Similar results have been
obtained using frequencies from GOLF.
These results agree with the results obtained using intermediate degree solar
p-modes \citep[e.g.,][]{ab94,pc94,ba95,basu98,rich98}.
This exercise therefore demonstrates that it is indeed
possible to determine helium abundance using only low degree modes.

\subsection[]{Stellar models beyond the main sequence}
\label{subsec:subg}

\begin{figure}
  \begin{center}
    \leavevmode
  \centerline{\epsfig{file=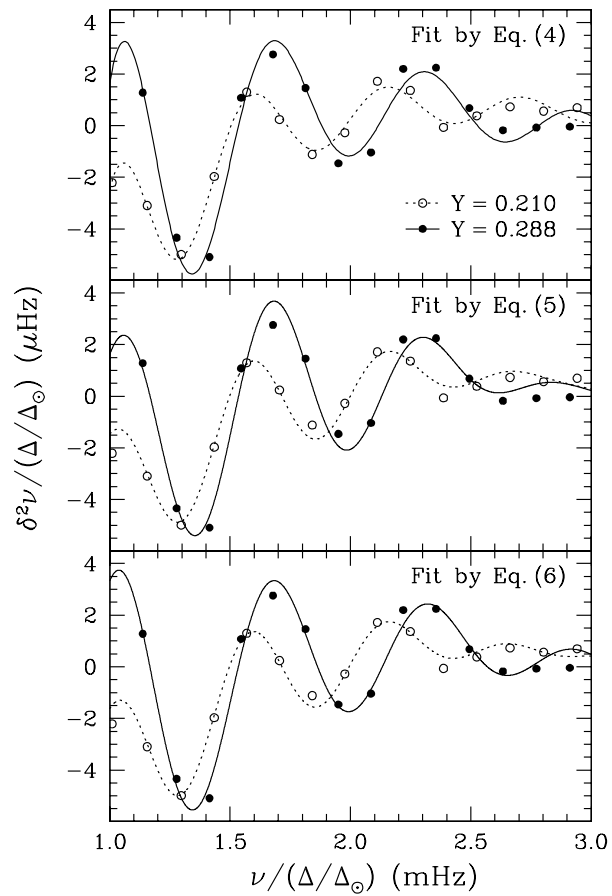,width=8.0cm}}
  \end{center}
\caption{A sample of the fits to the second differences of the
scaled frequencies using only $\ell=0$ modes for models evolved beyond
the main sequence.
The points are the `data', and the lines the
fits to the points. The examples shown are for $1.2M_\odot$ model evolved
to a radius of $1.8R_\odot$.}
\label{fig:subgfit}
\end{figure}

All  the stellar models that we have considered so far are on the
main sequence. We
have constructed a few models which have evolved beyond the
main sequence to see how we might determine the envelope helium
abundance for evolved stars.  For these stars,
many nonradial modes show a mixed character
where the same mode behaves like a gravity mode in the core and
like an acoustic mode in the envelope. The analysis that leads to the  calculation
of the oscillatory signal in frequencies is not valid for such modes.
In fact, even calculating the large frequency separation from the
calculated frequencies is somewhat difficult for such stars.
However,  radial modes are assured to be purely acoustic in nature 
throughout the star, and hence,   for such
stars we restrict our analysis to radial modes ($\ell=0$) only.
Thus the 
number of modes available for fitting is  small and we do not expect to
be able to calculate all parameters in Eqs.~(\ref{eq:quad}), (\ref{eq:mont}) or
(\ref{eq:quad2})  reliably. Since the amplitude
of the oscillatory signal due to the base of the convection zone is generally
much smaller than that due to the HeII ionisation zone, we only 
fit  the signal due to the HeII ionisation zone. Sample
fits are shown in Fig.~\ref{fig:subgfit}. The fits are reasonably
good, though not as good as those in Fig.~\ref{fig:diff}, hence we expect
somewhat larger systematic errors. Since the amplitude of the HeII signal
increases with reduction in frequency, we have extended the
fitting interval by including modes with frequencies 1--3~mHz. The
enhanced range increases the number of modes available as well as
the mean amplitude of the signal, thus providing better characterisation
of the signal.
When compared to main sequence stars, we expect larger uncertainties 
arising from errors in frequency because of the fewer number of modes
in these stars.
Nevertheless, Monte-Carlo simulations
with comparable error in frequencies show that the estimated error in
mean amplitude for evolved stars is not too much larger than  that for
main-sequence stars.

\begin{figure*}
  \begin{center}
    \leavevmode
  \centerline{\epsfig{file=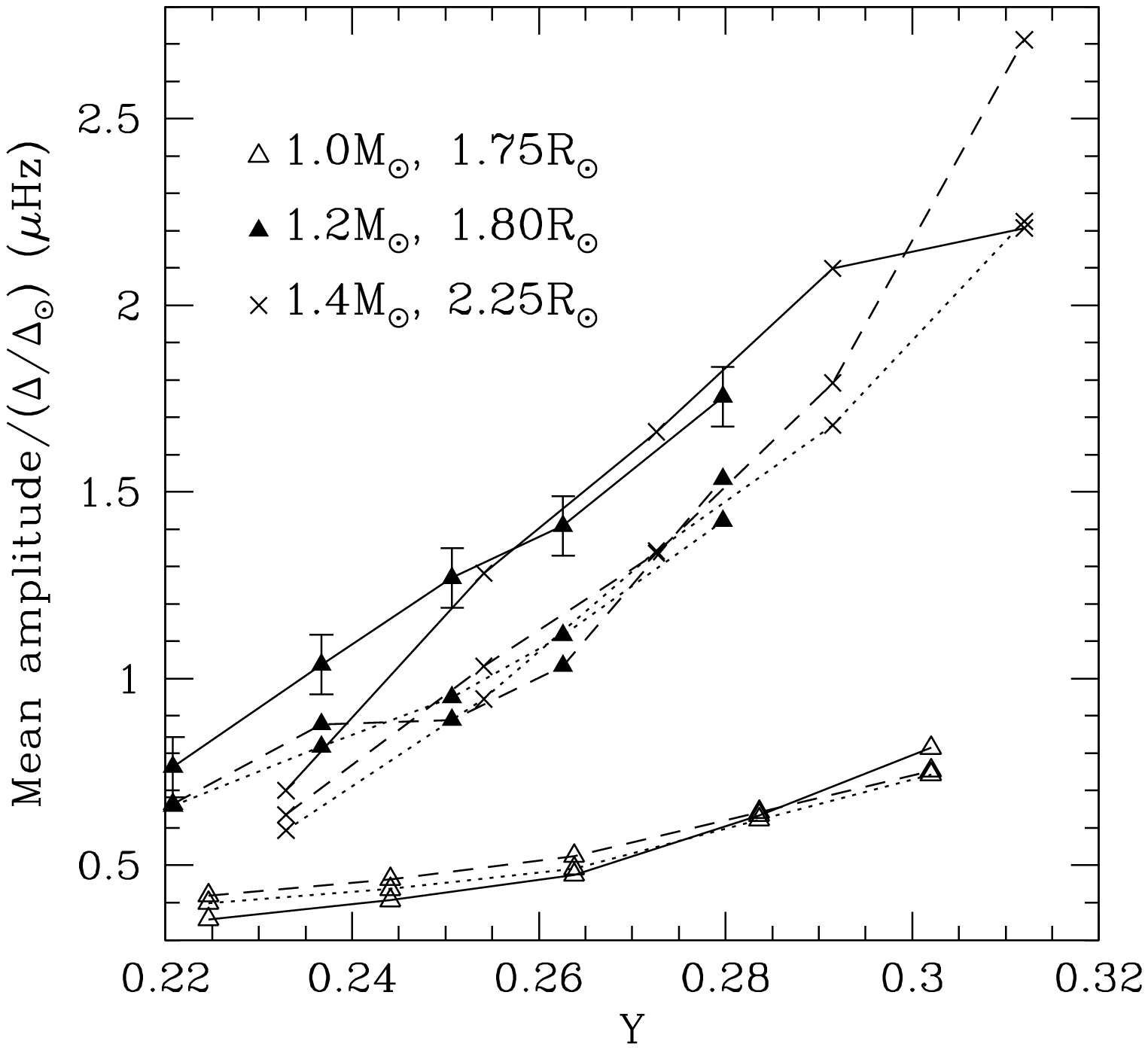,width=8.0cm}\hspace{1cm}
  \epsfig{file=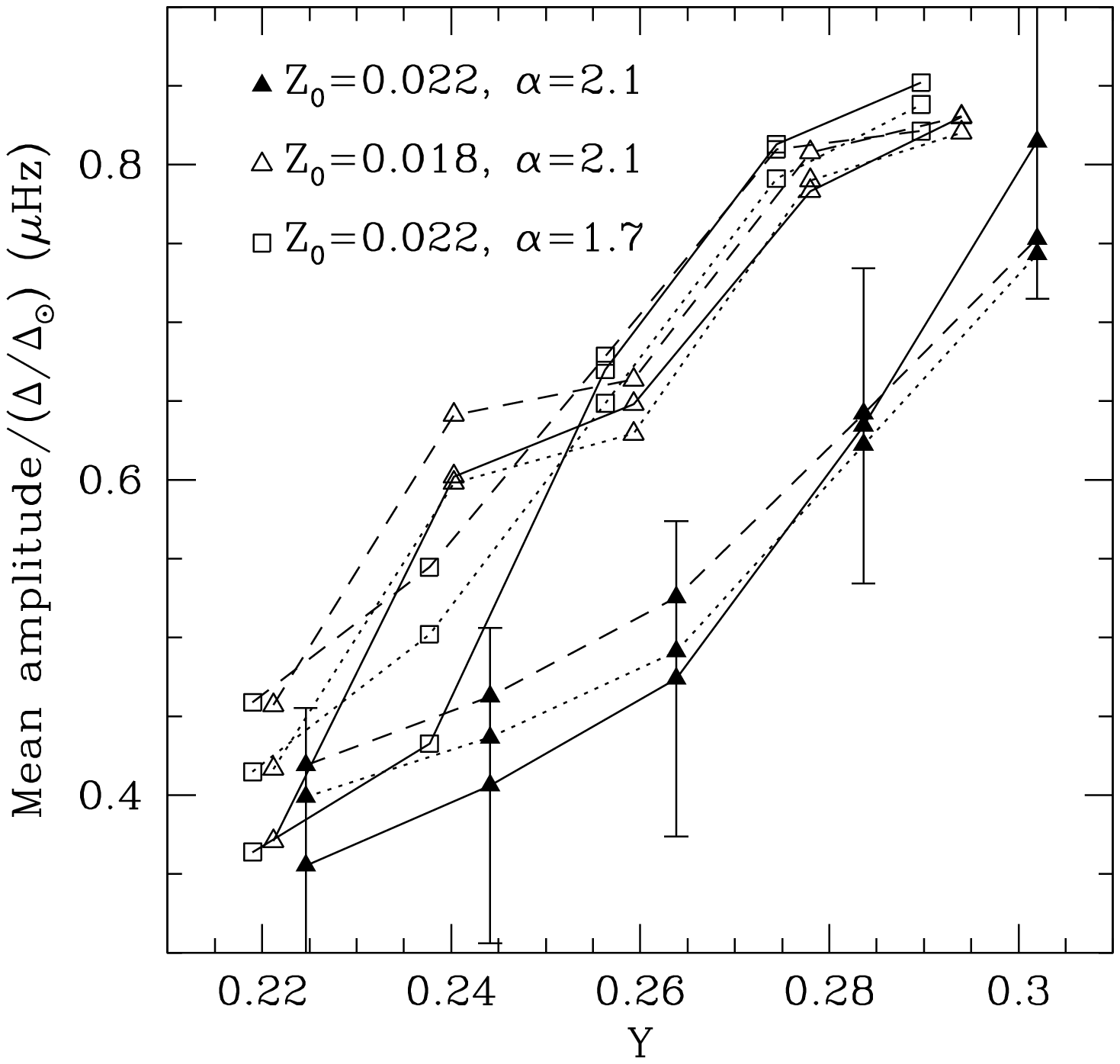,width=8.0cm}}
  \end{center}
  \caption{The mean amplitude
of the oscillatory part in frequency due to the HeII ionisation zone
is shown as a function of the helium abundance in the envelope for
evolved stellar models with a fixed mass and radius.
The solid, dotted and dashed lines
show the results for fits using Eqs.~(\ref{eq:quad}), (\ref{eq:mont})
and (\ref{eq:quad2}) respectively. The left panel shows the results
for a few different masses and radii as marked in the figure, while
the right panel shows the results for $M=1M_\odot$, $R=1.75R_\odot$
models with different $Z_0$ and $\alpha$. For sake of clarity, the
error estimates are shown on only one set of points.}
\label{fig:subgamp}
\end{figure*}

Figure~\ref{fig:subgamp} shows the amplitude of the signal as a function of $Y$ for
models with a fixed mass and radius. Error bars corresponding to frequency errors
of 1 part in $10^4$ are also shown for a few typical models.  
As expected the amplitude increases with $Y$
and it is possible to determine $Y$ from known amplitude. We find again that
these amplitudes are not particularly sensitive to small variations in $Z_0$
and $\alpha$. However, in this case the differences due to $Z_0$ and
$\alpha$ are larger than those for main sequence models.
As can be seen, despite the errors, we should be able to determine
the helium abundance in the envelope of these stars to a precision of 
about 0.01--0.02.

\subsection[]{Results for models with low heavy element abundances}
\label{ref:otherz}

Stars with very low heavy element abundances evolve quite differently 
from stars with near-solar heavy element abundances.  This is
because the convection zone of low $Z$ stars is
shallower at the same effective temperature \citep{ddk90,pm91}.
We do not expect this to affect the
helium signature; nevertheless, we confirm this using models of low $Z_0$.
We have constructed two sets of models, models with $Z_0=0.007$
and models with $Z_0=0.001$.
We have both main sequence and subgiant models with $Z_0=0.007$, but
only  subgiants with $Z_0=0.001$ since Population~II  models evolve very quickly compared
with stars of the same mass with higher $Z$ and
hence we do not expect to see many main-sequence stars.  As with the $Z_0=0.022$ models,
we use $\ell=0$,1,2 and 3 modes for the main sequence stars and only
$\ell=0$ modes for the subgiants.

\begin{figure*}
  \begin{center}
    \leavevmode
  \centerline{\epsfig{file=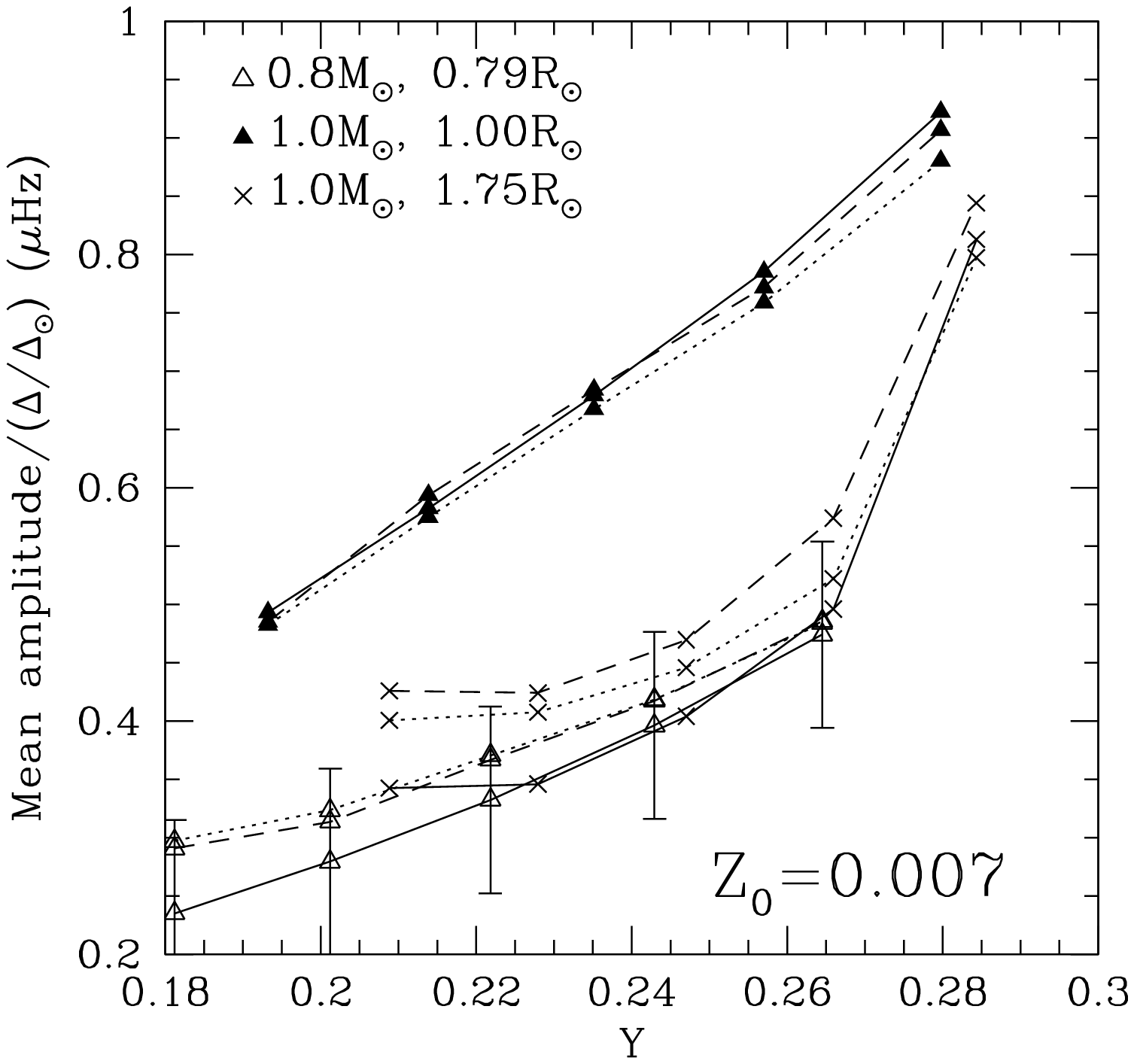,width=8.0cm}\hspace{1cm}
  \epsfig{file=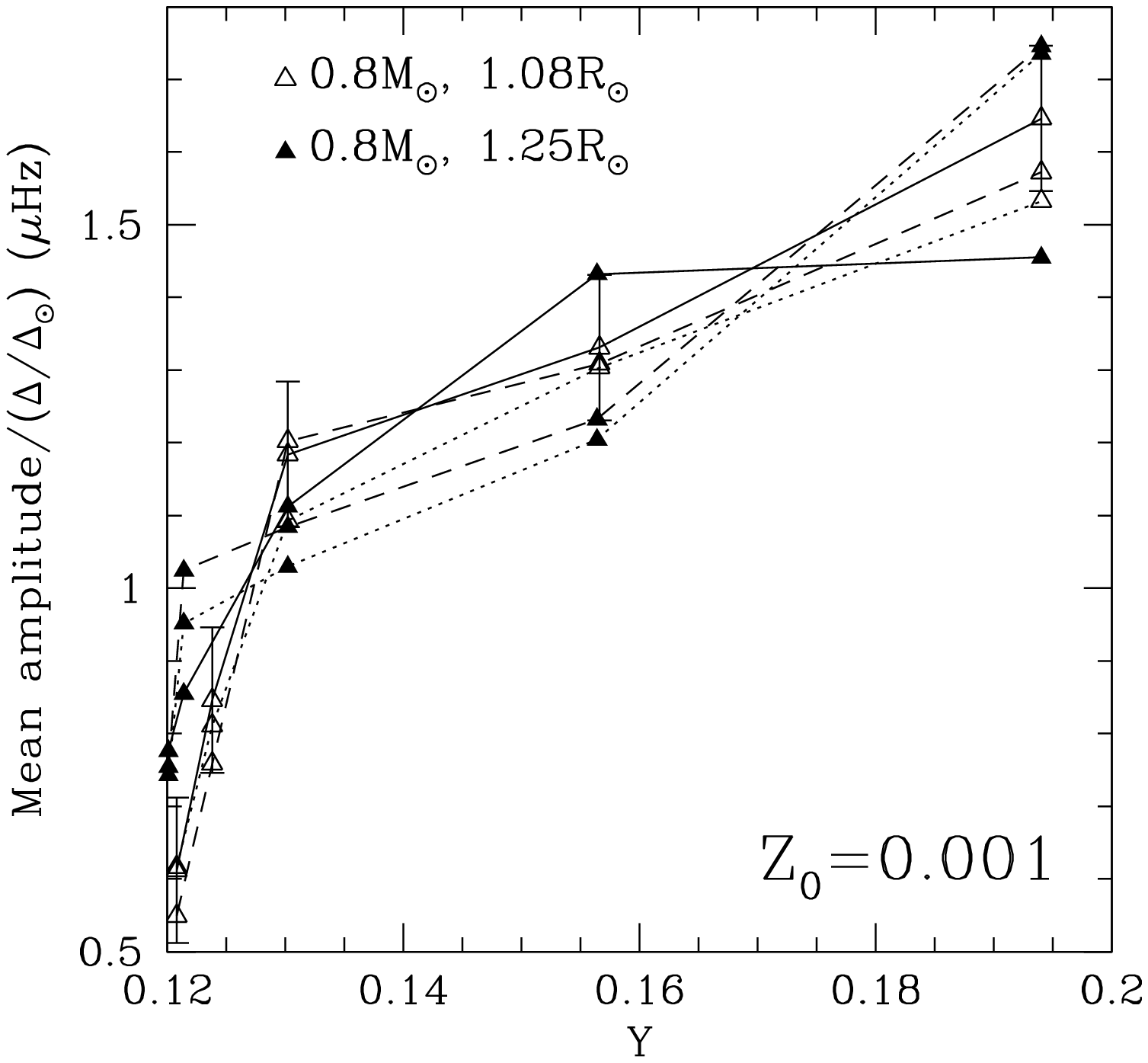,width=8.0cm}}
  \end{center}
  \caption{The mean amplitude
of the oscillatory part in frequency due to the HeII ionisation zone
is shown as a function of the helium abundance in the envelope for
stellar models with fixed mass and radius, constructed with
low heavy element abundance. This figure shows
the results for stellar models with initial low heavy element abundance,
$Z_0=0.007$ (left panel) and $Z_0=0.001$ (right panel).
The solid, dotted and dashed lines
show the results for fits using Eqs.~(\ref{eq:quad}), (\ref{eq:mont})
and (\ref{eq:quad2}) respectively. For sake of clarity, the
error estimates are shown on only one set of points.}
\label{fig:pop2}
\end{figure*}

Fig.~\ref{fig:pop2} shows the amplitude as a function of $Y$ for $Z_0=0.007$ models.
Again we can see that the amplitude increases with $Y$. The amplitudes are however
slightly different
from those of $Z_0=0.022$ models, and hence for such large differences in
$Z_0$ it will be better to make different calibration models.
Fig.~\ref{fig:pop2} also shows the results for Population~II models. Again the amplitude
increases with $Y$.
The amplitude for these evolved models are much larger than those for
main sequence models studied in Sect.~\ref{subsec:std}.
Nevertheless, to get good results for $Y$, we  would need lower errors in frequencies.
We expect to be able to observe only lower mass Population~II stars, and
this implies that the expected amplitude of the oscillation signal will be small, making errors larger.

\section[]{Conclusions}
\label{sec:conclusions}

We have presented results of our
investigation into how low degree oscillation frequencies may be
used to determine the helium abundance in stars with solar-type
oscillations.
We have demonstrated that in principle, it is possible to use this
technique to estimate the helium abundance in stellar envelopes using
the frequencies of low degree modes of oscillation.

We find that the oscillatory signal in the
frequencies due to the helium ionisation zone can be used to determine
the helium abundance of a low-mass main-sequence star, provided either the
radius or the mass is known independently.
The precision to which we may be able to
determine the helium abundance increases with increase in
mass.
Using reasonable error estimates it appears that a precision of
0.01--0.02 in $Y$ is possible in most cases.
The amplitude of the oscillatory term is not particularly sensitive to small
variations of the initial  heavy element abundance, $Z_0$ and mixing length
parameter, $\alpha$. For stars with masses of about  $1.4M_\odot$ and larger
it is difficult to fit the oscillatory signal reliably and hence
the uncertainties are larger. This difficulty is most probably
due to the fact that these stars have shallow convection zones and
there is not much difference in the acoustic depths of the HeII ionisation
zone and that of the base of the convection zone, thus leading to interference
between the two terms. For low mass stars the amplitude of the oscillatory
term is small leading to large errors in determining the helium abundance.
Thus the best results are obtained for stars with masses close to
and larger than the solar mass.
Using solar data, we have shown that we can indeed determine the
envelope helium abundance using low degree modes.

Even if the mass and radius for a star is not known independently,
we can use the large frequency separation $\Delta$ to estimate
$M/R^3$ and in that case also a reasonably tight correlation is
found between $Y$ and amplitude of the oscillatory signal, thus allowing
us to determine $Y$, though with larger errors.

For stellar models that are evolved off the main sequence it is not
possible to use  nonradial modes, however by using only radial modes
it is possible to fit the oscillatory term arising from the HeII
ionisation zone, which can then be used to measure the helium abundance.
A more detailed study of stars in different post main sequence phases
is required to understand the oscillatory signal from such stars.
Models with much lower heavy element abundances behave in a manner
similar to those with near-solar heavy element abundances as far as the
helium signature is concerned.

Although most of the present work uses exact frequencies from stellar models, 
we have successfully applied this method to solar low degree data.
We have also estimated the expected errors assuming frequency errors
of one part in $10^4$, a level that is expected to be achieved
by future space- and ground-based observations.
This opens up the possibility of using asteroseismic data to obtain
the helium abundance in stellar envelopes, leading to a better
understanding of the evolution of stellar populations in our galaxy. 

\section*{Acknowledgments}

This work utilises data from the Solar Oscillations
Investigation/ Michelson Doppler Imager (SOI/MDI) on the Solar
and Heliospheric Observatory (SOHO). SOHO is a project of
international cooperation between ESA and NASA.
MDI is supported by NASA grants NAG5-8878 and NAG5-10483
to Stanford University.
AM was supported in this work by CEFIPRA (Centre Franco-Indien pour la
Promotion de la Recherche Avanc\'e) project No.\ 2504-3.
PD was supported by NASA grant NAG5-13299.

\bsp

\label{lastpage}

\end{document}